\documentclass[%
superscriptaddress,
showpacs,
amsmath,amssymb,
twocolumn,
prcfloatfix,]
{revtex4-1}
\usepackage{color}
\definecolor{mygray}{gray}{.9}
\definecolor{mypink}{rgb}{.99,.91,.95}
\definecolor{mycyan}{cmyk}{.3,0,0,0}
\usepackage{graphicx}
\usepackage{dcolumn}
\usepackage{bm}
\usepackage{amsmath,amssymb}
\usepackage{booktabs}
\usepackage{natbib}

\usepackage{longtable}

\begin{document}
	
    \title{Influences of $\bm{Z=100}$ and $\bm{N=152}$ deformed shells on $\bm{ K^{\pi}=8^{-} }$ isomers and rotational bands in $\bm{N = 150}$ isotones}
 
    \author{Jun Zhang}
	\affiliation{College of Physics, Nanjing University of Aeronautics and Astronautics, Nanjing 210016, China}
	\author{Hai-Qian Zhang}
	\affiliation{College of Materials Science and Technology, Nanjing University of Aeronautics and Astronautics, Nanjing 210016, China}
    \author{T. M. Shneidman}
    \affiliation{Bogoliubov Laboratory of Theoretical Physics, Joint Institute for Nuclear Research, Dubna 141980, Russia}
    \affiliation{Kazan Federal University, Kazan 420008, Russia}
    \author{R. V. Jolos}
    \affiliation{Joint Institute for Nuclear Research, Dubna 141980, Russia}
    \affiliation{Dubna State University, Dubna 141982, Russia}
    \author{Xiao-Tao He}
	\email{hext@nuaa.edu.cn}
	\affiliation{College of Materials Science and Technology, Nanjing University of Aeronautics and Astronautics, Nanjing 210016, China}
    
	\date{\today}
\begin{abstract}
The $K^{\pi}=8^{-}$ isomeric states and rotational bands in the even-even $N = 150$ isotones with $94 \leqslant Z \leqslant 104$ are investigated by the cranked shell model (CSM) with pairing correlations treated by the particle-number-conserving (PNC) method. 
The experimental bandhead energies and kinematic moments of inertia (MOIs) are reproduced quite well by the PNC-CSM calculation. The two-neutron state with configuration $\nu 9/2^{-}[734] \otimes \nu 7/2^{+}[624]$ is the lowest $8^{-}$ state for these isomers. This is a demonstration of the deformed neutron shell at $N=152$. Low-lying two proton $\pi^{2}8^{-}$($\pi 9/2^{+}[624] \otimes \pi 7/2^{-}[514]$) configuration state is predicted only for $^{252}$No and $^{254}$Rf due to the deformed proton shell at $Z=100$. A distinct upbending is observed for the $\nu^{2}8^{-}$ bands in the lighter isotones while it is absent for bands in the heavier ones. The upbending of the $\nu^{2}8^{-}$ band at frequency $\hbar\omega\approx 0.20$ MeV in $^{244}$Pu attributes to the sudden proton alignment of the interference term $j_x(\pi5/2^{+}[642]\otimes\pi7/2^{+}[633])$. 
The irregularity of MOI observed in the $K^{\pi}=8^{-}$ band of $^{252}$No can be explained by the mixing of the $\nu^{2}8^{-}$($\nu 9/2^{-}[734] \otimes \nu 7/2^{+}[624]$) and $\pi^{2}8^{-}$($\pi 9/2^{+}[624] \otimes \pi 7/2^{-}[514]$) configurations. The $20\%-30\%$ increase of the bandhead $J^{(1)}$ for the $8^{-}$ bands comparing to the ground-state band is attributed to the $\sim 5\%$ pairing gap reduction of the two-neutron $\nu 7/2^{+}[624] \otimes \nu 9/2^{-}[734]$ configuration state comparing to the ground-state band.

\end{abstract}
	\maketitle

\section{Introduction}
Since the first discovery of isomeric states in $^{250}$Fm and $^{254}$No \cite{GhiorsoA1973_PRC7_2032}, many in-beam and decay spectroscopy studies have been performed on the light superheavy nuclei around the $Z=100$, $A=250$ mass region \cite{HotaS2016_PRC94_21303, ShirwadkarU2019_PRC100_34309, KatoriK2008_PRC78_14301, GreenleesP2008_PRC78_21303, SulignanoB2012_PRC86_44318, HessergerF2010_EPJA43_55, KhuyagbaatarJ2021_PRC103_64303, DavidH2015_PRL115_132502, TakahashiR2010_PRC81_57303, SeweryniakD2023_PRC107_61302, HerzbergR2008_PiPaNP61_674}. Besides the ground-state band, many multi-particle states have been observed in this mass region. The low-lying multi-particle states can provide valuable information on the nuclear structure, such as single-particle levels, pairing correlation, Pauli blocking effect and so on. In addition, the rotational bands built on these states can further provide insight into the nuclear rotational properties.

Two-particle $K^{\pi}=8^{-}$ isomers were systematically observed in the even-even $N=150$ isotones ranging from Pu ($Z=94$) to Rf ($Z=104$). These isomers are suggested to be associated with either a two-proton $\pi 9/2^{+}[624] \otimes \pi 7/2^{-}[514]$ or two-neutron $\nu 9/2^{-}[734] \otimes \nu 7/2^{+}[624]$ configuration~\cite{GhiorsoA1973_PRC7_2032,GreenleesP2008_PRC78_21303, TandelS2006_PRL97_82502, RobinsonA2008_PRC78_34308, ClarkR2010_PLB690_19, SulignanoB2012_PRC86_44318, KhuyagbaatarJ2020_NPA994_121662, JeppesenH2009_PRC79_31303, DavidH2015_PRL115_132502, KhuyagbaatarJ2021_PRC103_64303, RobinsonA2011_PRC83_64311, HotaS2016_PRC94_21303, ShirwadkarU2019_PRC100_34309, KatoriK2008_PRC78_14301, OrlandiR2022_PRC106_64301,MoodyK1987_ZPA328_417,SulignanoB2007_EPJA33_327,JolosR2011_JPG38_115103}. The rotational bands built on these isomers were identified in $^{244}$Pu \cite{HotaS2016_PRC94_21303}, $^{250}$Fm \cite{GreenleesP2008_PRC78_21303}, and $^{252}$No \cite{SulignanoB2012_PRC86_44318}. The observed $K^{\pi}=8^{-}$ band exhibited an upbending at rotational frequency $\hbar\omega\approx0.25$ MeV in $^{244}$Pu~\cite{HotaS2016_PRC94_21303}, while it remains nearly constant with increasing rotational  frequency in $^{250}$Fm \cite{GreenleesP2008_PRC78_21303}. As for $^{252}$No, the $K^{\pi}=8^{-}$ band shows an irregularity at rotational frequency $\hbar\omega\approx0.175$ MeV. The different rotational behaviors of these $K^{\pi}=8^{-}$ bands need further investigations and better understandings.\par

Theoretically, the isomeric $K^{\pi}=8^{-}$ states in this mass region have been investigated by Hartree-Fock-Bogoliubov (HFB) method with the D1S Gogny interaction \cite{HotaS2016_PRC94_21303}, the projected shell model \cite{HerzbergR2006_N442_896}, the cranked shell model \cite{HeX2020_CPC44_34106}, the Woods-Saxon (WS) method with the Lipkin-Nogami pairing formalism \cite{HotaS2016_PRC94_21303, TandelS2006_PRL97_82502, RobinsonA2008_PRC78_34308}, the quasiparticle phonon model \cite{JolosR2011_JPG38_115103}, and the configuration-constrained potential-energy-surface model \cite{LiuH2014_PRC89_44304, FuX2014_PRC89_54301}. As for the rotational behaviors of these nuclei, various theoretical calculations mainly focused on the properties of the ground-state bands \cite{DelarocheJ2006_NPA771_103, ZhangZ2012_PRC85_14324, ZhangZ2013_PRC87_54308, HeX2020_CPC44_34106, LiuH2012_PRC86_11301,HerzbergR2008_PiPaNP61_674}, with the exception of the study on the configuration mixing of the $K^{\pi}=8^{-}$ bands \cite{FuX2014_PRC89_54301}.
To our best knowledge, there is still no theoretical calculations systematically investigating the rotational behaviors of $K^{\pi}=8^{-}$ bands in the $N=150$ isotones.

The particle-number-conserving method in the framework of the cranked shell model (PNC-CSM) is one of the most useful models to investigate the low-lying high-$K$ rotational bands in the light superheavy nuclei. In the PNC-CSM method, the ground state and low-lying pair-broken excited states are obtained by diagonalizing the cranked shell model Hamiltonian in a truncated cranked many-particle configuration space directly, so that the particle number is conserved exactly and the Pauli blocking effects are taken into account simultaneously \cite{ZengJ1994_PRC50_746, ZengJ1983_NPA405_1}. 
The previous PNC-CSM calculation \cite{HeX2020_CPC44_34106} have well described the observed high-$K$ isomers and rotational bands in $^{254}$No \cite{TandelS2006_PRL97_82502, EeckhaudtS2005_EPJA26_227, EeckhaudtS2005_EPJA25_605, ReiterP1999_PRL82_509, ClarkR2010_PLB690_19}.\par

In this work, the PNC-CSM method is applied to investigate the $K^{\pi}=8^{-}$ isomers and rotational bands in the $N = 150$ isotones with $94 \leqslant Z \leqslant 104$. The configuration assignments of the observed $K^{\pi}=8^{-}$ isomers are confirmed. The different rotational behaviors observed in the $8^{-}$ rotational bands in $^{244}$Pu and $^{250}$Fm are discussed in details. The irregularity of the moment of inertia of $8^{-}$ band in $^{252}$No is explained.
The difference of the moments of inertia between the $K^{\pi}=8^{-}$ bands and ground-state bands is investigated.

\section{Theoretical framework}{\label{Sec:PNC}}
The cranked shell model Hamiltonian of an axially deformed nucleus in the rotating frame is
\begin{equation}
H_\mathrm{CSM}=H_\mathrm{0}+H_\mathrm{P}=H_{\rm Nil} -\omega J_{x} +H_\mathrm{P}(0)+H_\mathrm{P}(2),
\end{equation}
where $H_{\rm Nil}$ is the Nilsson Hamiltonian, and $-\omega J_{x}$ is the Coriolis interaction with cranking frequency $\omega$ about the $x$ axis (perpendicular to the nuclear symmetry $z$ axis). $H_{\rm P}$ is the pairing including monopole and quadrupole pairing correlations,
\begin{equation}
H_\mathrm{P}(0)= -G_{0} \sum_{\xi\eta}a_{\xi}^{\dagger}a_{\overline{\xi}}^{\dagger}a_{\overline{\eta}}a_{\eta},
\end{equation}
\begin{equation}
H_\mathrm{P}(2)= -G_{2} \sum_{\xi\eta}q_{2}(\xi)q_{2}(\eta)a_{\xi}^{\dagger}a_{\overline{\xi}}^{\dagger}a_{\overline{\eta}}a_{\eta},
\end{equation}
where $a_{\xi}^{\dagger}a_{\overline{\xi}}^{\dagger}$ ($a_{\overline{\eta}}a_{\eta}$) labels the pairing creation (annihilation) operator, $q_{2}(\xi) = \sqrt{{16\pi}/{5}}\langle \xi |r^{2}Y_{20}|\xi\rangle$ is the diagonal element of the stretched quadrupole operator, and the $G_{0}$ and $G_{2}$ are the effective strengths of monopole and quadrupole pairing interactions, respectively.\par

By diagonalizing $H_{\rm CSM}$ in a sufficiently large cranked many-particle configuration (CMPC) space, a sufficiently accurate low-lying excited eigenstates are obtained,
\begin{equation}
|\psi\rangle = \sum_{i}C_{i}|i\rangle ,
\label{eigenstate}
\end{equation}
where $C_{i}$ is real and $|i\rangle$ is a CMPC (an eigenstate of the one-body operator $H_{0}$) of the $n$-particle system. A detailed description of the diagonalization procedure is given in Ref. \cite{ZengJ1994_PRC50_1388}.\par

The occupation probability $n_{\mu}$ of the cranked orbital $|\mu\rangle$ can be calculated as
\begin{equation}
n_{\mu} = \sum_{i}|C_{i}|^{2} P_{i \mu} ,
\end{equation}
where $P_{i \mu} = 1$ if $|\mu\rangle$ is occupied in $|i\rangle$, and $P_{i \mu} = 0$ otherwise. The total particle number $N = \sum_{\mu} n_{\mu}$. By analyzing the $\omega$ dependence of the occupation probability $n_{\mu}$, the configuration of low-lying high-$K$ excited state and the upbending mechanism can be understood clearly.\par

The angular momentum alignment for the eigenstate $|\psi\rangle$ includes the diagonal and the off-diagonal parts,
\begin{equation}
\langle\psi|J_x|\psi\rangle = \sum_{i}{C_{i}}^{2}\langle i|J_x|i \rangle+ 2 \sum_{i < j}{C_{i}C_{j}}\langle i|J_x|j \rangle ,
\end{equation}
and $\langle\psi|J_x|\psi\rangle$ is denoted simply as $\langle J_x \rangle$ hereafter.\par 

Since $J_{x}$ is an one-body operator, the off-diagonal matrix element $\langle i|J_x|j \rangle$ ($i \neq j$) does not vanish only when $|i \rangle$ and $|j \rangle$ differ by one particle occupation. After a certain permutation of creation operators, $|i \rangle$ and $|j \rangle$ can be reconstructed into $|i \rangle = (-1)^{M_{i \mu}} |\mu \cdots\rangle, \quad  |j \rangle = (-1)^{M_{j \nu}} |\nu \cdots\rangle$, where $\mu$ and $\nu$ denote two different single-particle states and the ellipses stand for the same particle occupations. $(-1)^{M_{i \mu}} =\pm 1$, $(-1)^{M_{j \nu}} =\pm 1$ according to whether the permutation is even or odd. Thus, the angular momentum alignment $\langle \psi|J_x|\psi \rangle$ can be expressed as:
\begin{align}
\langle \psi|J_x|\psi \rangle &= \sum_{\mu} j_x (\mu) + \sum_{\mu < \nu} j_x{(\mu \nu)}, \nonumber \\
j_x (\mu) &= \langle \mu |j_x| \mu \rangle n_{\mu}, \nonumber \\
j_x (\mu\nu) &= 2 \langle \mu |j_x| \nu \rangle \sum_{i < j} (-1)^{M_{i \mu} + M_{j \nu}} C_{i}C_{j} \quad(\mu \neq \nu),
\label{diagonal and off-diagonal parts}
\end{align}
where $j_x (\mu)$ and $j_x{(\mu \nu)}$ are the diagonal and off-diagonal contribution, respectively.\par

The kinematic and dynamic moments of inertia for the eigenstate $|\psi\rangle$ are given by
\begin{equation}
J^{(1)}=\dfrac{\langle\psi|J_x|\psi\rangle}{\omega}, \quad \ J^{(2)}=\dfrac{d \langle\psi|J_x|\psi\rangle}{d \omega} . 
\end{equation}
\par

For the two-particle state in an even-even nucleus, $|i \rangle = |\sigma_{1}\sigma_{2}\mu_{1}\overline{\mu}_{1} \cdots\mu_{k}\overline{\mu}_{k} \rangle$, in which $\sigma_{1}$ and $\sigma_{2}$ are the Nilsson orbitals blocked by two unpaired particles. When the two unpaired particles are coupled, the parity $\pi = \pi_{\sigma_{1}}\pi_{\sigma_{2}}$, and the projections of their total angular momentum on the symmetry axis can produce two states with $K = | \Omega_{\sigma_{1}} \pm \Omega_{\sigma_{2}} |$. According to the Gallagher-Moszkowski (GM) rules \cite{GallagherC1962_PR126_1525}, the spin-singlet coupling is energetically favored for the isomeric $K^{\pi} = 8^{-}$ states in the $N = 150$ isotones studied in this work.\par

\begin{table}[!htp]
	\caption{Deformation parameters $\varepsilon_2$, $\varepsilon_4$, and $\varepsilon_6$ used in the present PNC-CSM calculations for the $N = 150$ isotones from $^{244}$Pu to $^{254}$Rf.}
	\setlength{\tabcolsep}{6.5mm}
	\renewcommand{\arraystretch}{1.25}
	{
		\begin{tabular}{l c c c }
			\hline\hline
			
			& $\varepsilon_2$ & $\varepsilon_4$  & $\varepsilon_6$  \\[3pt]
			\hline
			
			$^{244}$Pu   &0.260  &-0.010  &0.042\\[3pt]
			$^{246}$Cm   &0.250  &-0.015  &0.035\\[3pt]
			$^{248}$Cf   &0.250  &0.015   &0.042\\[3pt]
			$^{250}$Fm   &0.255  &-0.010   &0.040\\[3pt]
			$^{252}$No   &0.260  &0.030   &0.030\\[3pt]
			$^{254}$Rf   &0.260  &0.035   &0.035\\[3pt]
			
			\hline\hline
		\end{tabular}
	}
	\label{deformation parameters}
\end{table}
  
\begin{figure}[!htp]
	\includegraphics[scale=0.34]{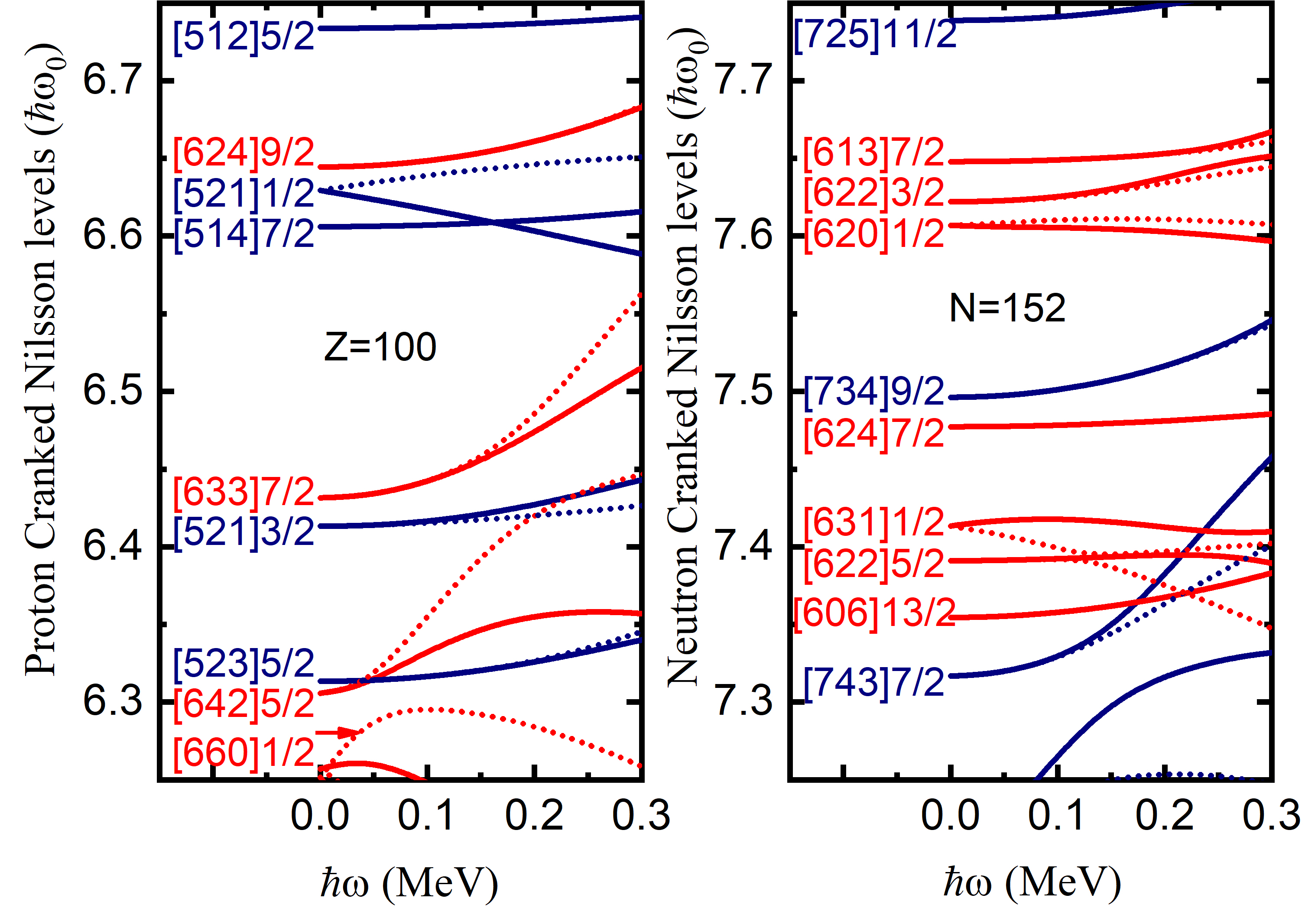}
	\caption{(Color online) The cranked Nilsson levels near the Fermi surface of $^{248}$Cf for protons (left) and neutrons (right). The positive (negative) parity levels are denoted by red (blue) lines. The signature $\alpha = +1/2$ ($\alpha = -1/2$) levels are denoted by solid (dotted) lines.}
	\label{Cf248_SingleParticlelevels}
\end{figure}
 
\section{RESULTS AND DISCUSSIONS}{\label{Sec:Results}}
\subsection{Cranked Nilsson levels}
The Nilsson parameters ($\kappa, \mu$) proposed in Ref. \cite{ZhangZ2012_PRC85_14324} are optimized to reproduce the experimental single-particle levels of the odd-$A$ nuclei in the actinide mass region, which are obtained with only deformation $\varepsilon_2$ and $\varepsilon_4$. Studies show that high-order deformation $\varepsilon_6$ plays an important role on the excitation energies and the kinematic moments of inertia for the superheavy nuclei \cite{LiuH2011_PRC83_11303, PatykZ1991_NPA533_132, HeX2020_CPC44_34106, ZhangZ2018_PRC98_34304}. Thus, the values of proton $\kappa_5,\mu_5$ and neutron $\kappa_6,\mu_6$ are modified slightly to reproduce the single-particle level sequence with the addition of high-order deformation $\varepsilon_6$ in the present work.\par 

The deformations are input parameters in the PNC-CSM calculations, which are chosen to be close to the avaliable experimental values \cite{BastinJ2006_PRC73_24308,HerzbergR2001_PRC65_14303} and change smoothly according to the proton number.
The deformation parameters $\varepsilon_2$, $\varepsilon_4$, and $\varepsilon_6$ used in the present PNC-CSM calculations for the $N = 150$ isotones from $^{244}$Pu to $^{254}$Rf are listed in Table \ref{deformation parameters}. The effective pairing strengths $G_{0}$ and $G_{2}$, in principle, can be determined by the odd-even differences in the experimental nuclear binding energies. The values are connected with the dimensions of the truncated CMPC space. In the present PNC-CSM calculations, the CMPC spaces for all nuclei are constructed in the proton $N=5,6$ shells and the neutron $N=6,7$ shells. The dimensions of the CMPC space are about 1000 for both protons and neutrons. The corresponding effective pairing strengths are $G_{0p} = 0.25$ MeV, $G_{2p} = 0.03$ MeV and $G_{0n} = 0.25$ MeV, $G_{2n} = 0.02$ MeV for protons and neutrons, respectively. The stability of the PNC-CSM calculations against the change of the dimensions of the CMPC space has been investigated in Refs. \cite{ZengJ1994_PRC50_746, LiuS2002_PRC66_67301}.\par

\begin{figure}[!htp]
	\includegraphics[scale=0.35]{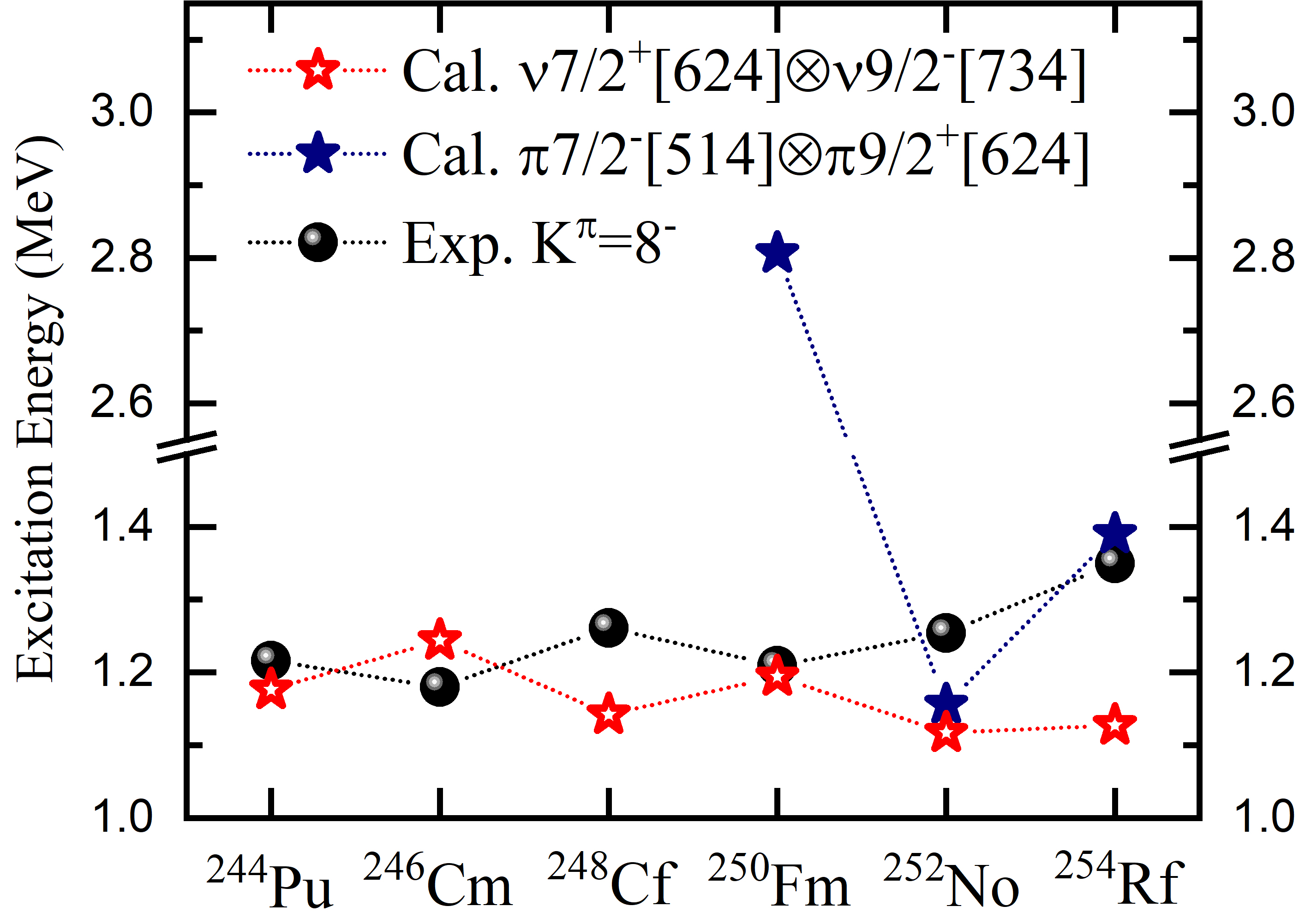}
	\caption{(Color online) Comparison of experimentally observed and calculated excitation energies of $K^{\pi}=8^{-}$ isomers in the $N = 150 $ isotones from $^{244}$Pu to $^{254}$Rf. The experimental data are taken from Refs. \cite{HotaS2016_PRC94_21303, ShirwadkarU2019_PRC100_34309, KatoriK2008_PRC78_14301, GreenleesP2008_PRC78_21303, SulignanoB2012_PRC86_44318, DavidH2015_PRL115_132502}.}
	\label{BandheadEnergy_8-}
\end{figure}

Figure \ref{Cf248_SingleParticlelevels} shows the calculated cranked Nilsson levels near the Fermi surface of $^{248}$Cf. The positive (negative) parity levels are denoted by red (blue) lines. The signature $\alpha = + 1/2$ ($\alpha = - 1/2$) levels are denoted by solid (dotted) lines. Based on such a sequence of single-particle levels, the experimental one-particle states in the neighbor odd-$A$ nuclei are reproduced quite well, such as the one-proton states in $^{247}$Bk and $^{249}$Es, and one-neutron excited states in $^{247}$Cf and $^{249}$Cf with the only exception of the first excited state $\nu 5/2^{+}[622]$. The disagreement with the experimental result maybe caused by the residual octupole interaction between the $\nu 5/2^{+}[622]$ and $\nu 9/2^{-}[734]$ orbitals, which leads to a lower excitation energy of the $\nu 5/2^{+}[622]$ state \cite{HerzbergR2008_PiPaNP61_674, YatesS1975_PRC12_442}. Octupole correlations are ignored in the present calculations.\par

As shown in Fig. \ref{Cf248_SingleParticlelevels}, there exist a proton gap at $Z = 100$ and a neutron gap at $N = 152$, which are consistent with the calculation by the Woods-Saxon potential \cite{ChasmanR1977_RMP49_833}. For $^{244}$Pu, $^{246}$Cm, and $^{248}$Cf, the proton high-$j$ orbitals around the Fermi surfaces are proton $\pi 1/2^{+}[660]$, $\pi 5/2^{+}[642]$, and $\pi 7/2^{+}[633]$ levels stemming from the $\pi i_{13/2}$ orbitals.\par

\begin{figure*}[!htp]
	\includegraphics[scale=0.35]{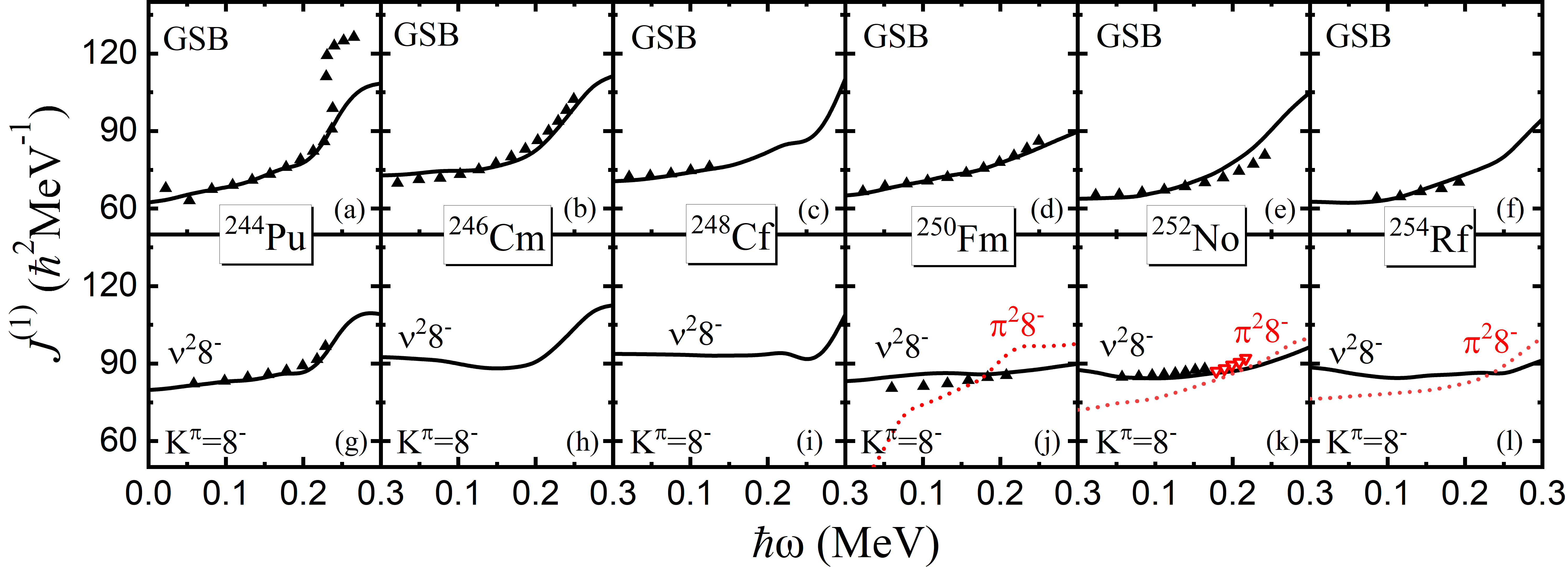}
	\caption{(Color online) The experimental (symbols) and calculated (lines) kinematic moments of inertia $J^{(1)}$ for the ground-state bands and the isomeric $K^{\pi} = 8^{-}$ bands in the $N = 150$ isotones from $^{244}$Pu to $^{254}$Rf. The available experimental data are taken from Refs. \cite{HotaS2016_PRC94_21303, HerzbergR2008_PiPaNP61_674, TakahashiR2010_PRC81_57303, GreenleesP2008_PRC78_21303, SulignanoB2012_PRC86_44318, SeweryniakD2023_PRC107_61302}. The $\nu^{2} 8^{-}$ stands for the two-neutron $\nu 7/2^{+}[624] \otimes \nu 9/2^{-}[734]$ configuration, and $\pi^{2} 8^{-}$ stands for the two-proton $\pi 7/2^{-}[514] \otimes \pi 9/2^{+}[624]$ configuration.}
	\label{MOI_gsb&8-}
\end{figure*}

\subsection{$\bm{K^{\pi} = 8^{-}}$ isomeric states}
The $K^{\pi} = 8^{-}$ isomers are observed systematically in the $N = 150$ isotones $^{244}$Pu \cite{HotaS2016_PRC94_21303}, $^{246}$Cm \cite{ShirwadkarU2019_PRC100_34309}, $^{248}$Cf \cite{KatoriK2008_PRC78_14301}, $^{250}$Fm \cite{GreenleesP2008_PRC78_21303}, $^{252}$No \cite{SulignanoB2012_PRC86_44318}, and $^{254}$Rf \cite{DavidH2015_PRL115_132502}. The comparison of the excitation energies between the PNC-CSM calculations and the experimental data are shown in Fig. \ref{BandheadEnergy_8-}. The experimental data are denoted by black solid balls, the calculated two-neutron $8^{-}$ states with the configuration $\nu 7/2^{+}[624] \otimes \nu 9/2^{-}[734]$ are denoted by red open stars, and the calculated two-proton $8^{-}$ states with the configuration $\pi 7/2^{-}[514] \otimes \pi 9/2^{+}[624]$ are denoted by blue solid stars. 

Experimentally, the $K^{\pi}=8^{-}$ isomers in the $N=150$ isotones, from $^{244}$Pu to $^{252}$No, are located very close to each other in excitation energies with $E \approx 1.20$ MeV, while the energy of $8^{-}$ isomer is higher at about 1.35 MeV in $^{254}$Rf \cite{DavidH2015_PRL115_132502}. 

In the present PNC-CSM calculations, the two-neutron $8^{-}$ states can well reproduce the experimental data in $^{244}$Pu, $^{246}$Cm, $^{248}$Cf, and $^{250}$Fm, and the deviation of excitation energies are less than 0.05 MeV. Thus, the $K^{\pi}=8^{-}$ isomers in $^{244}$Pu, $^{246}$Cm, $^{248}$Cf, and $^{250}$Fm are assigned unambiguously as a two-neutron $8^{-}$ state with the configuration $\nu 7/2^{+}[624] \otimes \nu 9/2^{-}[734]$. For $^{250}$Fm, besides the two-neutron $8^{-}$ isomer, the present calculation also predict a two-proton $8^{-}$ state with the configuration $\pi 7/2^{-}[514] \otimes \pi 9/2^{+}[624]$ at 2.806 MeV. However, the energy of the two-proton $8^{-}$ state is too high to be the observed isomer. 

For $^{252}$No, since the Fermi surface is above the $Z=100$ deformed shell gap, two low-lying $8^{-}$ states with similar excitation energies are predicted. One is a two-proton $8^{-}$ state with the configuration $\pi 7/2^{-}[514] \otimes \pi 9/2^{+}[624]$, and the other is a two-neutron $8^{-}$ state with the configuration $\nu 7/2^{+}[624] \otimes \nu 9/2^{-}[734]$. The experimental excitation energy can be reproduced quite well by both of the configurations, thus the configuration of the $K^{\pi}=8^{-}$ isomer in $^{252}$No needs further study of the rotational behavior, which will be discussed in section \ref{Sec:Irregularity}. 

In $^{254}$Rf, the two-neutron $8^{-}$ state with the configuration $\nu 7/2^{+}[624] \otimes \nu 9/2^{-}[734]$ is the lowest lying $8^{-}$ state, which is in consistent with the calculations in Refs. \cite{DavidH2015_PRL115_132502, LiuH2014_PRC89_44304}. However, it locates too low to reproduce the experimental data. Whereas, the two-proton $8^{-}$ state with the configuration $\pi 7/2^{-}[514] \otimes \pi 9/2^{+}[624]$ at the energy 1.390 MeV is more closer to the experimental data in the present PNC-CSM calculations.
In addition, the observed $25/2^{+}$ state in $^{255}$Rf \cite{ChakmaR2023_PRC107_14326} and $27/2^{+}$ state in $^{257}$Rf \cite{QianJ2009_PRC79_64319} are interpreted as a three-particle high-$K$ isomer originating from a $9/2^{-}[734]$ neutron configuration and a $11/2^{-}[725]$ neutron configuration coupled to the two-proton $\pi7/2^{-}[514]\otimes\pi9/2^{+}[624]$ configuration. Therefore, the possibility of two-proton $\pi7/2^{-}[514]\otimes\pi9/2^{+}[624]$ configuration cannot be ruled out for the $K^{\pi}=8^{-}$ isomer in $^{254}$Rf. 

The systematic presence of two-neutron $8^{-}$ state with the $\nu9/2^{-}[734]$ configuration coupling with $\nu7/2^{+}[624]$ configuration instead of $\nu7/2^{+}[613]$ at a low energy of $\approx1.20$ MeV in the $N=150$ isotones demonstrates the deformed neutron shell gap at $N=152$. Whereas, the calculated two-proton $8^{-}$ state in $N=150$ isotones with low energy can be obtained only for nuclei with proton number larger than $100$, i.e. $^{252}$No ($Z=102$) and $^{254}$Rf ($Z=104$). It is evidence that there is a significant proton shell gap at $Z=100$. In the lighter $N=150$ isotones, to form a two-proton $8^{-}$ state, the unpaired proton is required to be excited across the gap, which is reflected in obviously higher excitation energies.


\subsection{Moments of inertia}
Figure \ref{MOI_gsb&8-} shows the experimental (symbols) and calculated (lines) kinematic moments of inertia of the ground-state bands and the isomeric $K^{\pi} = 8^{-}$ rotational bands in the $N = 150$ isotones from $^{244}$Pu to $^{254}$Rf.
The experimental kinematic MOIs for each band are determined by $J^{(1)}(I)=(2I+1)/ E_{\gamma}(I+1 \rightarrow I-1)$.

The ground-state bands are systematically observed in all the isotones studied in this work \cite{HotaS2016_PRC94_21303, HerzbergR2008_PiPaNP61_674, TakahashiR2010_PRC81_57303, GreenleesP2008_PRC78_21303, SulignanoB2012_PRC86_44318, SeweryniakD2023_PRC107_61302}. From Fig. \ref{MOI_gsb&8-}(a), we can see that 
the experimental $J^{(1)}$ of GSB in $^{244}$Pu exhibits a sharp upbending at frequency $\hbar\omega \approx $ 0.25 MeV, which is distinctly different from those in the other isotones.
In Fig. \ref{MOI_gsb&8-}(b) there is just a slight upbending at $\hbar\omega \approx $ 0.20 MeV for the GSB in $^{246}$Cm. In Fig. \ref{MOI_gsb&8-}(c), the GSB in $^{248}$Cf is only observed at frequency $\hbar\omega \leqslant$ 0.15 MeV, and no upbending phenomenon is seen at low frequency region. Unlike the lighter isotones, from Figs. \ref{MOI_gsb&8-}(d)-\ref{MOI_gsb&8-}(f), we can see that only smooth increases of experimental MOIs in the GSBs are observed at the whole frequency region for the heavier isotones $^{250}$Fm, $^{252}$No, and $^{254}$Rf.

In Figs. \ref{MOI_gsb&8-}(a) and \ref{MOI_gsb&8-}(b), it can be seen that the upbending frequencies of calculated $J^{(1)}$ of GSB's in $^{244}$Pu and $^{246}$Cm are about 0.20 $\sim$ 0.25 MeV, which can well reproduce the experimental data. Furthermore, the present calculation predicts a slight upbending in $^{248}$Cf at $\hbar\omega\approx$ 0.25 MeV. It is well known that the backbending is caused by the crossing of the  GSB with a pair-broken band based on the high-$j$ intruder orbitals. In this mass region, the high-$j$ intruder orbitals involved near the Fermi surface are proton $\pi i_{13/2}$ and neutron $j_{15/2}$ orbitals. For the isotones $^{250}$Fm, $^{252}$No, and $^{254}$Rf, the gradual rise of the calculated $J^{(1)}$ are in consistent with the experimental data.

The rotational bands built on $K^{\pi}=8^{-}$ isomeric states were observed in the $N=150$ isotones $^{244}$Pu, $^{250}$Fm, and $^{252}$No \cite{HotaS2016_PRC94_21303,GreenleesP2008_PRC78_21303,SulignanoB2012_PRC86_44318}. 
From Fig. \ref{MOI_gsb&8-}(g), we can see that the experimental $J^{(1)}$ of the isomeric $K^{\pi}=8^{-}$ band in $^{244}$Pu parallels that of the ground-state band, and exhibits a upbending at the same frequency $\hbar\omega\approx$ 0.25 MeV. While, in Fig. \ref{MOI_gsb&8-}(j), the rotational behavior is much plainer for the isomeric band in $^{250}$Fm. It can be seen from Fig. \ref{MOI_gsb&8-}(k) that the experimental MOI of isomeric band in $^{252}$No exhibits a shallow minimum at frequency $\hbar\omega\approx$ 0.17 MeV.

In Figs. \ref{MOI_gsb&8-}(g)-\ref{MOI_gsb&8-}(l), the present PNC-CSM calculations present a two-neutron $8^{-}$ band in all the $N=150$ isotones studied in this work. $J^{(1)}$ of the two-neutron $8^{-}$ band with the configuration $\nu7/2^{+}[624]\otimes\nu9/2^{-}[734]$ shows upbendings at the frequency region $\hbar\omega=0.20\sim 0.25$ MeV in the lighter isotones $^{244}$Pu, $^{246}$Cm, and $^{248}$Cf, while it keeps nearly constant during the whole frequency region in the heavier isotones $^{250}$Fm, $^{252}$No, and $^{254}$Rf. Since the proton Fermi surfaces are different for these isotones, the contributions from protons depend strongly on the detailed single-particle level structure, such as high-$j$ intruder orbitals and the deformed shells.
As shown in Figs. \ref{MOI_gsb&8-}(j)-\ref{MOI_gsb&8-}(l), the present PNC-CSM calculations also predict the $8^{-}$ band with two-proton $\pi7/2^{-}[514]\otimes\pi9/2^{+}[624]$ configuration in $^{250}$Fm, $^{252}$No, and $^{254}$Rf. We can see that the $J^{(1)}$ of the two-proton $8^{-}$ band shows a smooth increase with frequency, while the two-neutron $8^{-}$ band is much plainer at the whole frequency region in these nuclei. 
For $^{250}$Fm, the excitation energy of the two-proton $8^{-}$ state can not agree well with the experimental data, which is larger than the experimental data by about 1.6 MeV [see Fig. \ref{BandheadEnergy_8-}]. Besides, the calculated $J^{(1)}$ of the two-neutron $8^{-}$ band well reproduce the experimental MOI, it is reasonable to consider that the $K^{\pi}=8^{-}$ band is a pure two-neutron $\nu7/2^{+}[624]\otimes\nu9/2^{-}[734]$ configuration at the whole rotation frequency region. 


\subsection{Different rotational behaviors between $\bm{^{244}}$Pu and $\bm{^{250}}$Fm}

\begin{figure}[!htp]
	\includegraphics[scale=0.32]{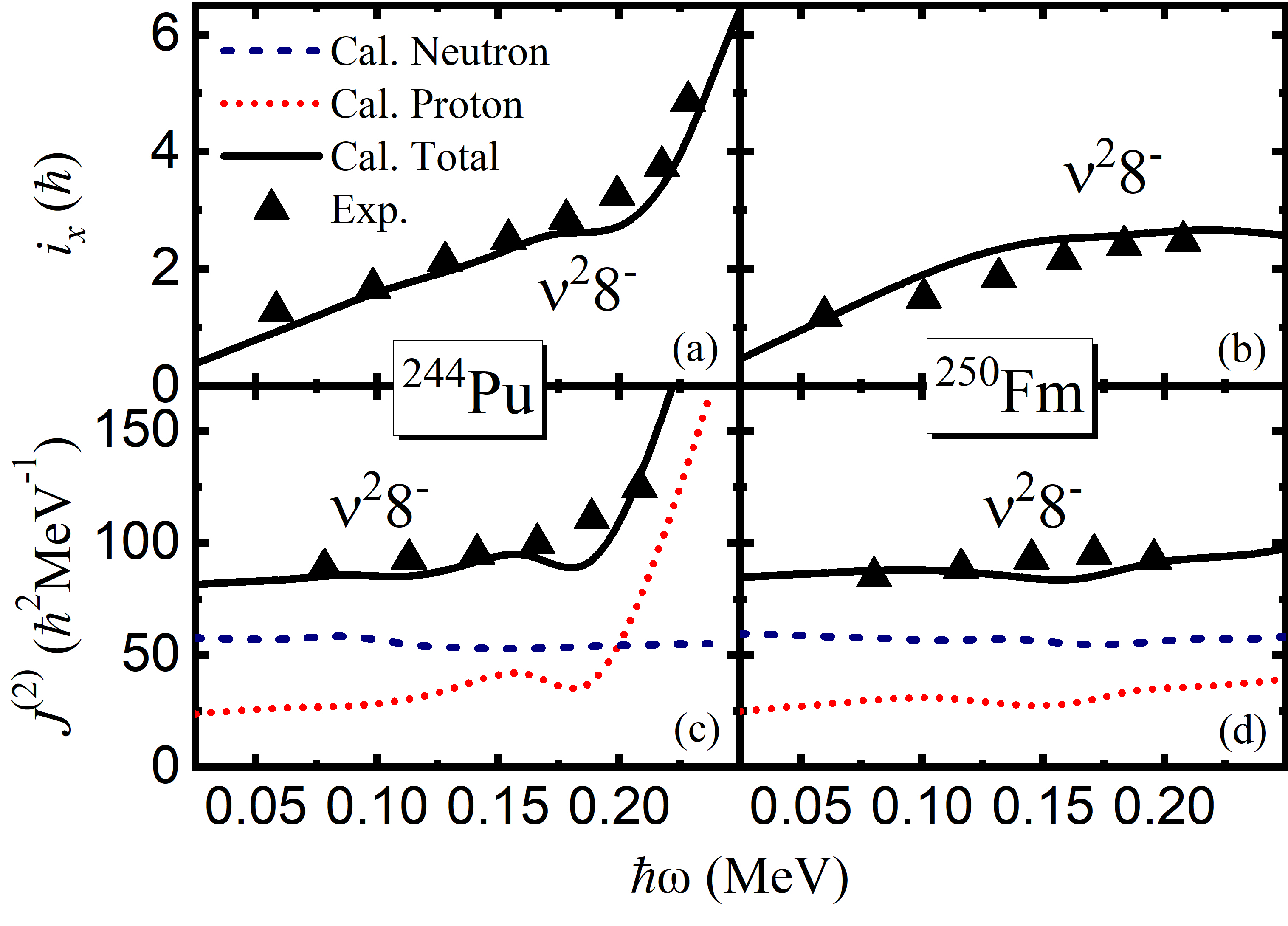}
	\caption{(Color online) The experimental (symbols) and calculated (lines) alignments $i_{x} = \langle J_{x} \rangle - \omega J_{0} - \omega^{3} J_{1}$ of the isomeric $K^{\pi} = 8^{-}$ bands with the two-neutron $\nu 7/2^{+}[624] \otimes \nu 9/2^{-}[734]$ configuration in the $N = 150$ isotones $^{244}$Pu and $^{250}$Fm. The Harris parameters $J_{0}$ = 65$\hbar ^{2}$ MeV$^{-1}$ and $J_{1}$ = 200$\hbar ^{4}$ MeV$^{-3}$. The available experimental data are taken from Refs. \cite{HotaS2016_PRC94_21303, GreenleesP2008_PRC78_21303}.}
	\label{Anignment_8-}
\end{figure}

The distinct rotational behaviors of the two-neutron $8^{-}$ bands between lighter isotones (including $^{244}$Pu, $^{246}$Cm, and $^{248}$Cf) and heavier isotones (including $^{250}$Fm, $^{252}$No, and $^{254}$Rf) are discussed in detail by selecting examples of $^{244}$Pu and $^{250}$Fm.
The experimental and calculated alignments and dynamic MOIs for the $K^{\pi} = 8^{-}$ bands with two-neutron $\nu 7/2^{+}[624] \otimes \nu 9/2^{-}[734]$ configuration in $^{244}$Pu and $^{250}$Fm are shown in Fig. \ref{Anignment_8-}. The experimental dynamic MOIs for each band are determined by $J^{(2)}(I)=4/[{E_{\gamma}(I+2 \rightarrow I)-E_{\gamma}(I \rightarrow I-2)}]$. The experimental data and calculated results are denoted by symbols and lines, respectively. It can be seen that the calculated results are in good agreement with the experimental data. In Fig. \ref{Anignment_8-}(a), the calculated upbending of $K^{\pi} = 8^{-}$ band occurs at frequency $\hbar\omega \approx 0.20$ MeV in $^{244}$Pu. In contrast, the alignment of $K^{\pi} = 8^{-}$ band in $^{250}$Fm increases slowly at $\hbar\omega < 0.20$ MeV, and keeps nearly constant at $\hbar\omega > 0.20$ MeV, as shown in Fig. \ref{Anignment_8-}(b). Moreover, as shown in Figs. \ref{Anignment_8-}(c) and \ref{Anignment_8-}(d), the calculated $J^{(2)}$ shows that the contribution from neutrons (blue dashed lines) are almost constant versus the frequency $\hbar\omega$ for both $^{244}$Pu and $^{250}$Fm, the different behavior of these two-neutron $8^{-}$ bands are mainly due to the contribution of protons (red dotted lines).\par

\begin{figure}[!htp]
	\includegraphics[scale=0.33]{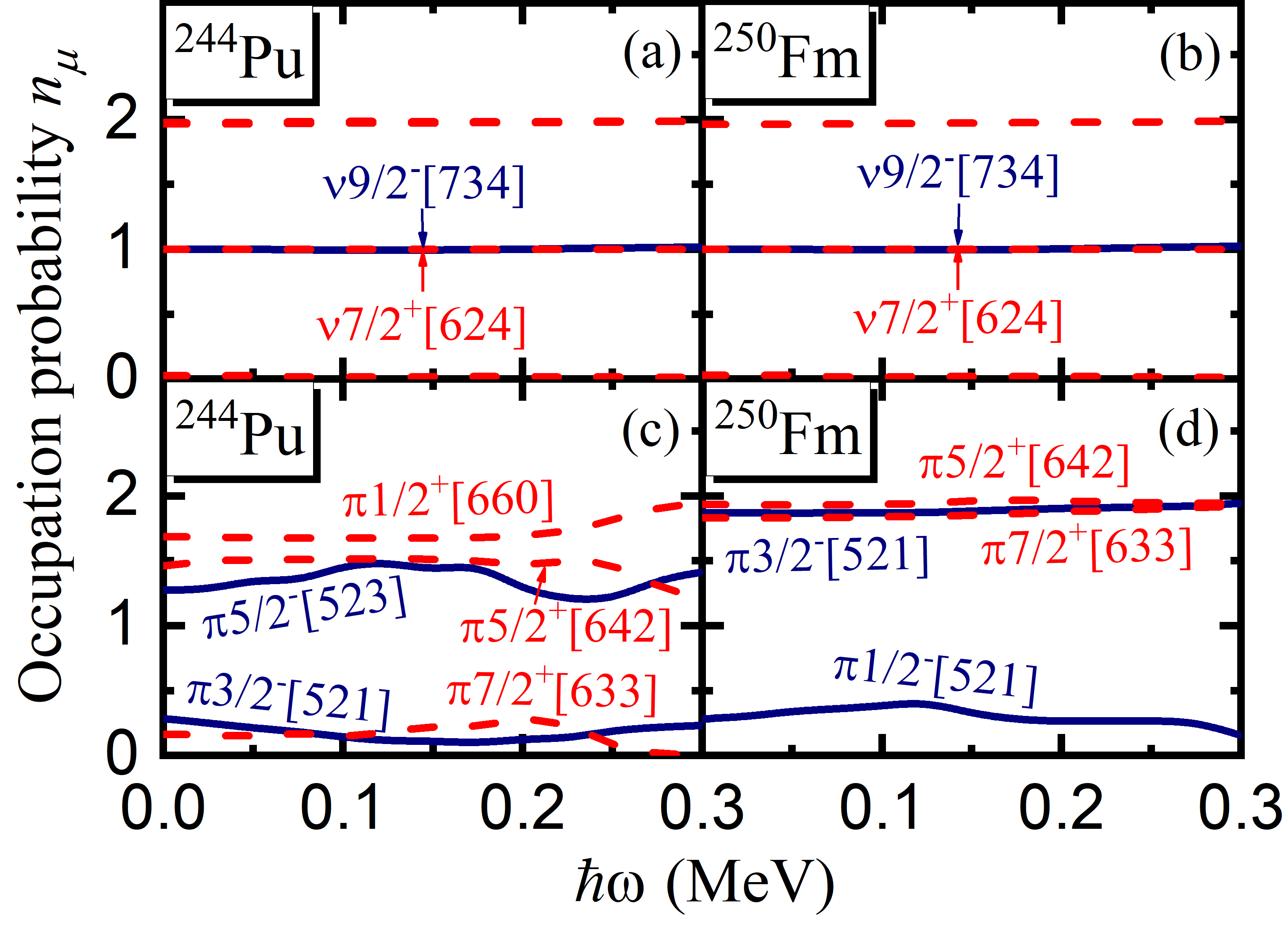}
	\caption{(Color online) Occupation probability $n_{\mu}$ of each orbital $\mu$ (including both $\alpha = \pm 1/2$) near the Fermi surface for the isomeric $K^{\pi} = 8^{-}$ bands for the two-neutron configuration $\nu 7/2^{+}[624] \otimes \nu 9/2^{-}[734]$ in the $N = 150$ isotones $^{244}$Pu and $^{250}$Fm. The top and bottom rows are for neutrons and protons, respectively. The positive and negative parity levels are denoted by red dashed and blue solid lines, respectively. The Nilsson levels far above the Fermi surface ($n_{\mu} \approx 0$) and far below ($n_{\mu} \approx 2$) are not shown.}
	\label{Occupation_8-}
\end{figure}

Figure \ref{Occupation_8-} shows the occupation probability $n_{\mu}$ of each cranked Nilsson orbital $\mu$ (including both $\alpha = \pm 1/2$) near the Fermi surface for the $K^{\pi} = 8^{-}$ bands in $^{244}$Pu and $^{250}$Fm. The top and bottom rows are for neutrons and protons, respectively. The positive (negative) parity levels are denoted by red dashed (blue solid) lines. The Nilsson levels far above the Fermi surface ($n_{\mu} \approx 0$) and far below ($n_{\mu} \approx 2$) are not shown. From Figs. \ref{Occupation_8-}(a) and \ref{Occupation_8-}(b), we can see that both of neutron orbitals $\nu 9/2^{-}[734]$ and $\nu 7/2^{+}[624]$ are blocked ($n_{\mu}$ $\approx$ 1) at the whole frequency region. Thus, the isomeric $K^{\pi}=8^{-}$ bands in $^{244}$Pu and $^{250}$Fm are both assigned as a two-neutron $\nu 7/2^{+}[624] \otimes \nu 9/2^{-}[734]$ configuration, which is consistent with the experimentally observed two-neutron bands.\par

For $^{244}$Pu, Fig. \ref{Occupation_8-}(c) shows that the high-$j$ orbitals $\pi i_{13/2}$ (including $\pi 1/2^{+}[660]$ and $\pi 5/2^{+}[642]$) and normal orbital $\pi 5/2^{-}[523]$ are partially occupied. This is because the Fermi surface of $^{244}$Pu locates far below the $Z=100$ deformed shell, the proton single-particle level density near the Fermi surface is quite high [see Fig. \ref{Cf248_SingleParticlelevels}]. Due to pairing correlations, orbitals above and below the Fermi surface are partially occupied.\par

Contrary to $^{244}$Pu, it can be seen in Fig. \ref{Occupation_8-}(d) that the occupation probabilities of proton orbitals in $^{250}$Fm are comparatively pure, i.e., either fully occupied ($n_{\mu}\approx 2$) or almost empty ($n_{\mu}\approx 0$). This is because the proton Fermi surface of $^{250}$Fm is just below the $Z=100$ proton deformed shell [see Fig. \ref{Cf248_SingleParticlelevels}].\par

\begin{figure}[!htp]
	\includegraphics[scale=0.31]{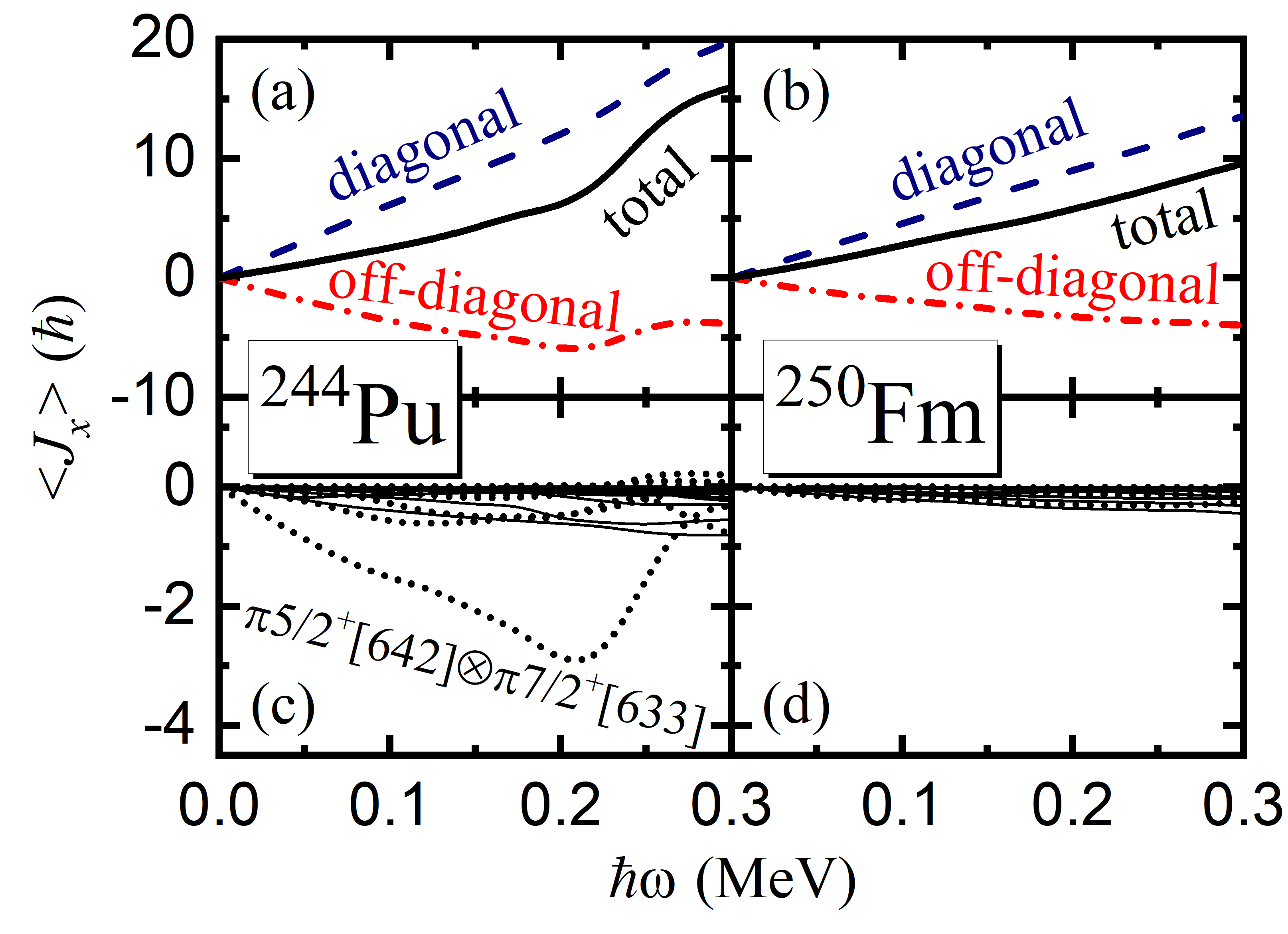}
	\caption{(Color online) Contributions of protons to the angular momentum alignments $\langle J_{x} \rangle$ for the isomeric $K^{\pi} = 8^{-}$ bands for the two-neutron configuration $\nu 7/2^{+}[624] \otimes \nu 9/2^{-}[734]$ in the $N = 150$ isotones $^{244}$Pu and $^{250}$Fm. The diagonal $\sum_{\mu} j_{x}(\mu)$ and off-diagonal parts $\sum_{\mu<\nu} j_{x}(\mu \nu)$ in Eq.(\ref{diagonal and off-diagonal parts}) are denoted by blue dashed and red dash-dotted lines, respectively (top row). The interference terms $j_{x}(\mu \nu)$ between orbitals stemming from high-$j$ intruder orbitals $\pi i_{13/2}$ are denoted by black dotted lines, other interference terms are denoted by black solid lines (bottom row).}
	\label{Jx_total_single}
\end{figure}

In order to have a more clear understanding of upbending mechanism, the contributions of protons to the angular momentum alignments $\langle J_{x} \rangle$ for the isomeric $K^{\pi} = 8^{-}$ bands in the $N = 150$ isotones $^{244}$Pu and $^{250}$Fm are shown in Fig. \ref{Jx_total_single}. The diagonal $\sum_{\mu} j_{x}(\mu)$ and off-diagonal parts $\sum_{\mu<\nu} j_{x}(\mu \nu)$ in Eq.(\ref{diagonal and off-diagonal parts}) are denoted by blue dashed and red dash-dotted lines, respectively (top row). From Fig. \ref{Jx_total_single}(a), it can be seen clearly that the upbending of proton alignments in $^{244}$Pu is mainly due to the off-diagonal contributions from protons. In Fig. \ref{Jx_total_single}(b), both diagonal and off-diagonal parts change gradually with the cranking frequency $\hbar\omega$ increasing, which result in the gradual rise of total alignment in $^{250}$Fm.\par 

Figs. \ref{Jx_total_single}(c) and \ref{Jx_total_single}(d) show the contribution of each proton orbital to the angular momentum alignments $\langle J_{x} \rangle$ for the $K^{\pi} = 8^{-}$ bands in the $N = 150$ isotones $^{244}$Pu and $^{250}$Fm, respectively. The interference terms $j_{x}(\mu \nu)$ between orbitals stemming from high-$j$ intruder orbitals $\pi i_{13/2}$ are denoted by black dotted lines, other interference terms are denoted by black solid lines. From Fig. \ref{Jx_total_single}(c), it can be seen that the upbending of alignment in the isomeric $K^{\pi} = 8^{-}$ band in $^{244}$Pu is mainly due to the off-diagonal part $j_{x} (\pi 5/2^{+}[642] \pi 7/2^{+}[633])$. However, as shown in Fig. \ref{Jx_total_single}(d), there are no upbendings for all the interference terms in $^{250}$Fm. \par
\subsection{Irregularity of moment of inertia of the isomeric $\bm{K^{\pi}=8^{-}}$ band in $\bm{^{252}}$No}{\label{Sec:Irregularity}

In $^{252}$No, the observed $K^{\pi} = 8^{-}$ rotational band based on the 1254-keV isomer exhibits irregularity in the moment of inertia at rotational frequency $\hbar\omega \approx 0.17$ MeV reported in Ref. \cite{SulignanoB2012_PRC86_44318}. It was tentatively interpreted as a crossing between the $K^{\pi} = 8^{-}$ band and $K^{\pi} = 2^{-}$ octupole band \cite{SulignanoB2012_PRC86_44318}. While Fu $et$ $al$. suggested that the MOI irregularity is caused by the configuration mixing with a two-proton $K^{\pi}=7^{-}$ band within a configuration-constrained total-Routhian-surface method \cite{FuX2014_PRC89_54301}.\par

\begin{figure}[!htp]
	\includegraphics[scale=0.32]{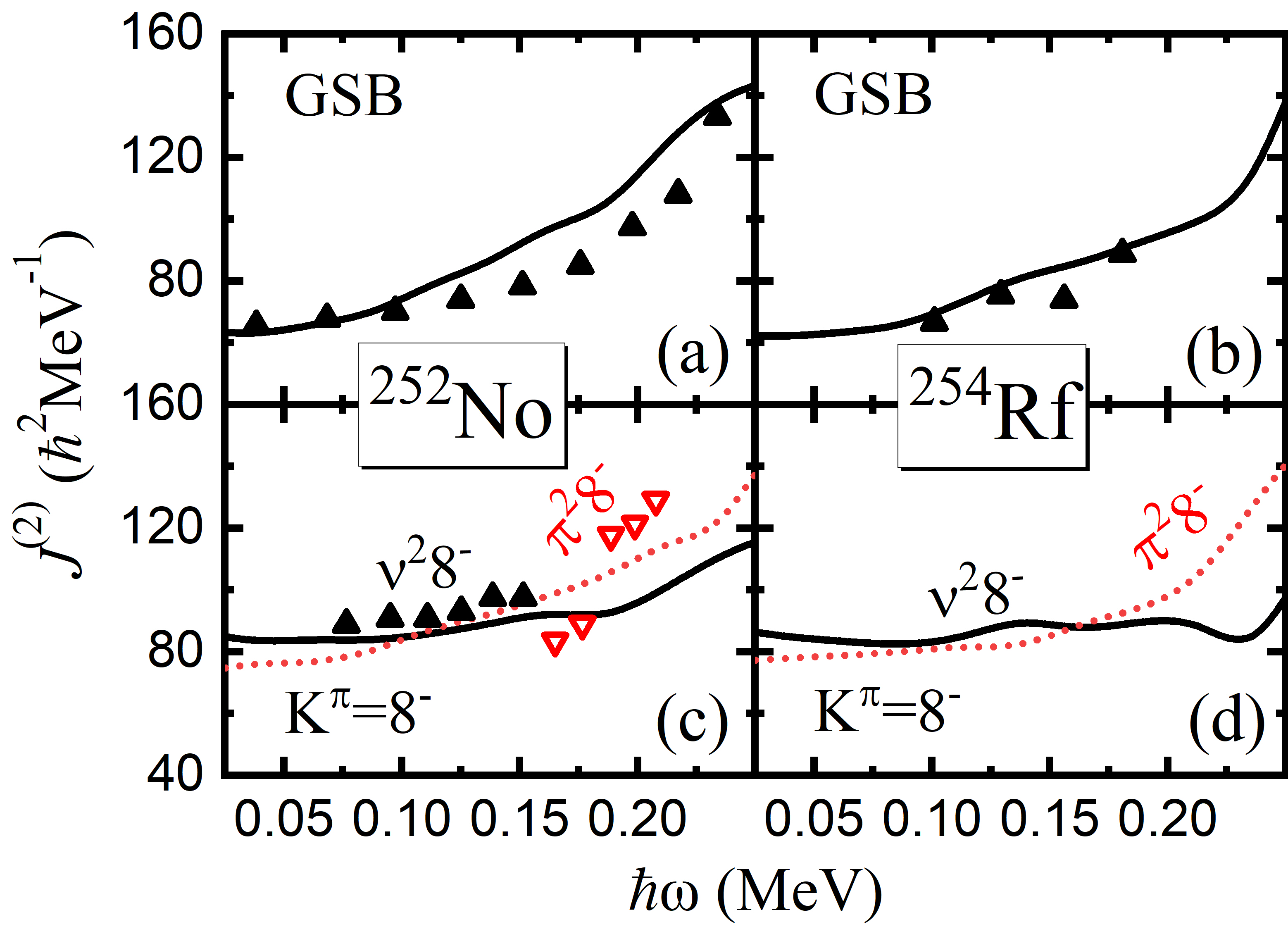}
	\caption{(Color online) The experimental (symbols) and calculated (lines) dynamic moments of inertia $J^{(2)}$ of the ground-state bands and the isomeric $K^{\pi} = 8^{-}$ bands in the $N = 150$ isotones $^{252}$No and $^{254}$Rf. The experimental data are taken from Ref. \cite{SulignanoB2012_PRC86_44318}.}
	\label{No252Rf254_J(2)_gsb&8-}
\end{figure}

Figure \ref{No252Rf254_J(2)_gsb&8-} shows the experimental and calculated dynamic moments of inertia $J^{(2)}$ for the ground-state bands and isomeric $K^{\pi} = 8^{-}$ bands in $^{252}$No and $^{254}$Rf. The experimental data are denoted by symbols, and the PNC-CSM calculations are denoted by lines. As shown in Figs. \ref{No252Rf254_J(2)_gsb&8-}(a) and \ref{No252Rf254_J(2)_gsb&8-}(b), the experimental $J^{(2)}$ of GSBs in both $^{252}$No and $^{254}$Rf are well reproduced by the PNC-CSM calculations.\par

In Fig. \ref{No252Rf254_J(2)_gsb&8-}(c), the PNC-CSM calculation give a two-neutron $8^{-}$ band with the configuration $\nu 7/2^{+}[624] \otimes \nu 9/2^{-}[734]$ and a two-proton $8^{-}$ band with the configuration $\pi 7/2^{-}[514] \otimes \pi 9/2^{+}[624]$ for the observed $K^{\pi}=8^{-}$ band in $^{252}$No.
Since the excitation energy of the two-neutron state is lower than that of the two-proton state [see Fig. \ref{BandheadEnergy_8-}], and the experimental $J^{(2)}$ is well reproduced by the calculation with the two-neutron configuration at frequency $\hbar\omega < 0.17$ MeV, the main component of the $K^{\pi} = 8^{-}$ band is expected to be the two-neutron $\nu 7/2^{+}[624] \otimes \nu 9/2^{-}[734]$ configuration at low rotational frequency. However, at the higher frequency ($\hbar\omega \geqslant 0.17$ MeV), the calculated $J^{(2)}$ with the two-proton $\pi 7/2^{-}[514] \otimes \pi 9/2^{+}[624]$ configuration agrees better with the experimental data. Thus, the MOI irregularity observed in the $K^{\pi} = 8^{-}$ band in $^{252}$No can be explained by a two-neutron configuration mixing with a two-proton configuration at frequency $\hbar\omega \approx 0.17$ MeV.\par

Similar situation happens in the $K^{\pi} = 8^{-}$ band in $^{254}$Rf, the two-neutron $\nu 7/2^{+}[624] \otimes \nu 9/2^{-}[734]$ configuration band crosses with two-proton $\pi 7/2^{-}[514] \otimes \pi 9/2^{+}[624]$ configuration band at $\hbar\omega \approx 0.15$ MeV in the present calculations, as shown in Fig. \ref{No252Rf254_J(2)_gsb&8-}(d). Due to the lack of experimental data for the isomeric $K^{\pi} = 8^{-}$ band in $^{254}$Rf, whether or not the irregularity of MOI occurs in the rotational band needs further experiments.\par 

\subsection{Pairing correlations in $\bm{^{252}}$No}
The bandhead of $J^{(1)}$ for the $8^{-}$ bands is larger than the ground-state bands by about $20\%-30\%$ in the $N=150$ isotones, which can be explained by pairing reduction of these bands due to the Pauli blocking effect.
In the PNC-CSM formalism, the nuclear pairing gap is defined as \cite{WuX2011_PRC83_34323}
\begin{equation}
  \tilde{\Delta}= G_{0}[-\frac{1}{G_{0}}\langle\psi|H_{\rm P}|\psi\rangle],  
\end{equation}
where $|\psi\rangle$ is a eigenstate [Eq. (\ref{eigenstate})] of the cranked shell model Hamiltonian $H_{\mathrm{CSM}}$.


\begin{figure}[!htp]
	\includegraphics[scale=0.26]{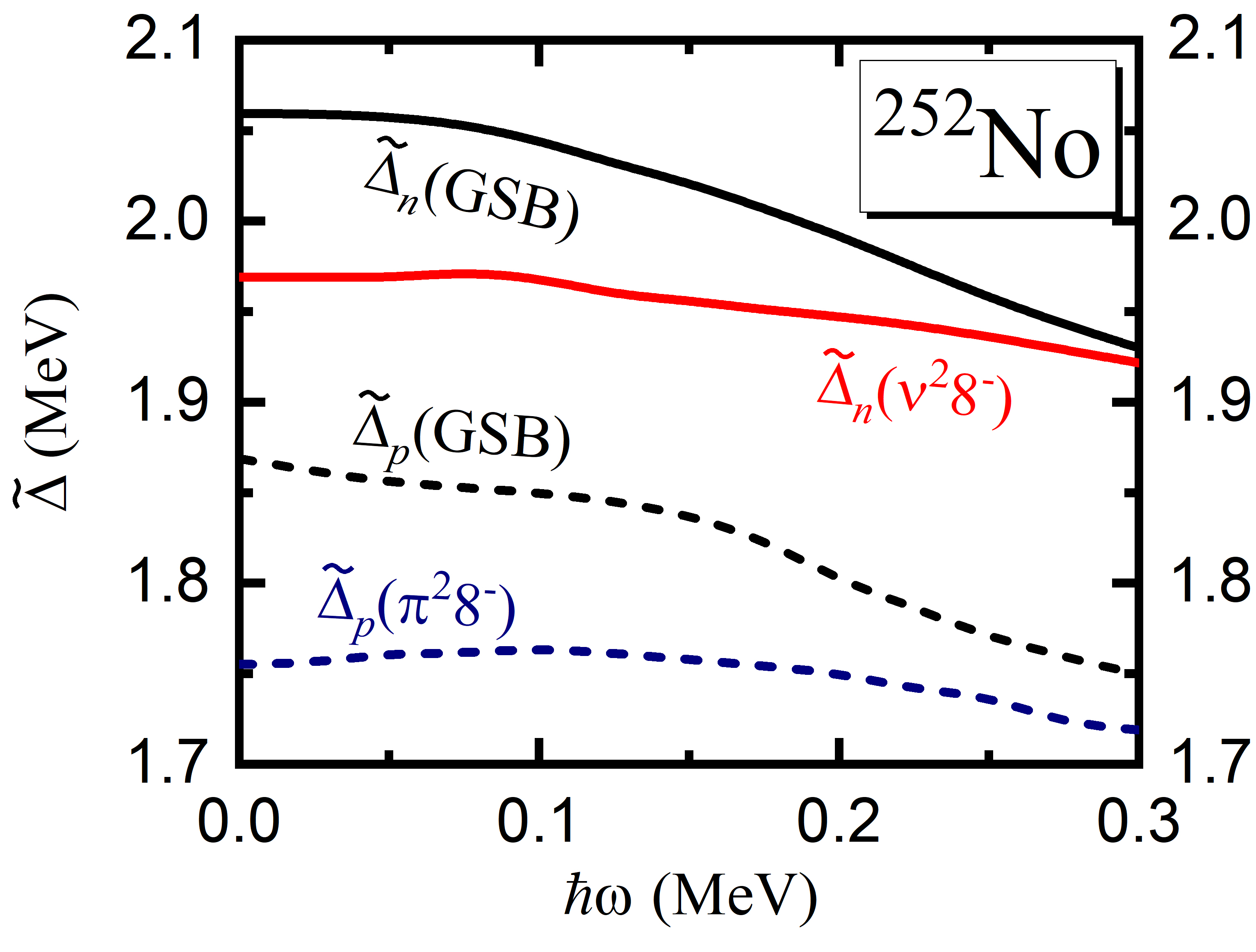}
	\caption{(Color online) The calculated pairing gaps $\tilde{\Delta}$ for the ground-state band (black lines), the two-neutron $8^{-}$ band with the configuration $\nu 7/2^{+}[624] \otimes \nu 9/2^{-}[734]$ (red line), and the two-proton $8^{-}$ band with the configuration $\pi 7/2^{-}[514] \otimes \pi 9/2^{+}[624]$ (blue line) in $^{252}$No.}
   \label{pairing}
\end{figure}

We take $^{252}$No for example to discuss. Figure \ref{pairing} shows the calculated $\tilde{\Delta}_n$ (neutrons) and $\tilde{\Delta}_p$ (protons) pairing gap for the ground-state, two-neutron $8^{-}$, and two-proton $8^{-}$ bands in $^{252}$No. The effective pairing strengths in the calculations are the same for the GSB and $8^{-}$ bands, the difference in the pairing gaps is purely from the wave functions. Both rotational frequency $\omega$ dependence and seniority $\nu$ (number of unpaired particles) dependence of the pairing gaps $\tilde{\Delta}$ are investigated. 

As a function of rotational frequency $\omega$, in general, the pairing gap
$\tilde{\Delta}(\omega)$ gradually decreases with increasing $\omega$. For the GSB, two-neutron, and two-proton $8^{-}$ bands, the PNC-CSM calculations for the pairing gap reduction with increasing $\omega$ are
\begin{equation}
\begin{aligned}
  \frac{\tilde{\Delta}_p(\hbar\omega=0.3\,\mathrm{MeV})-\tilde{\Delta}_p(\hbar\omega=0)}{\tilde{\Delta}_p(\hbar\omega=0)} & \approx 6.4\%,\quad \rm{GSB},  \\
\frac{\tilde{\Delta}_n(\hbar\omega=0.3\,\mathrm{MeV})-\tilde{\Delta}_n(\hbar\omega=0)}{\tilde{\Delta}_n(\hbar\omega=0)} & \approx 6.3\%,\quad \rm{GSB},  \\  
  \frac{\tilde{\Delta}_n(\hbar\omega=0.3\,\mathrm{MeV})-\tilde{\Delta}_n(\hbar\omega=0)}{\tilde{\Delta}_n(\hbar\omega=0)} & \approx 2.4\%,\quad \nu^{2}8^{-}, \\
  \frac{\tilde{\Delta}_p(\hbar\omega=0.3\,\mathrm{MeV})-\tilde{\Delta}_p(\hbar\omega=0)}{\tilde{\Delta}_p(\hbar\omega=0)} & \approx 2.1\%,\quad \pi^{2}8^{-}.     
\end{aligned}
\end{equation}
The frequency dependences of pairing gaps decrease about 6\% for GSB, and about 2\% (2\%) for the two-neutron (two-proton) $8^{-}$ band. Therefore, $J^{(1)}$ of the GSB displays an increase with the rotational frequency, while the $8^{-}$ band keeps almost flat with frequency in $^{252}$No [see Figs. \ref{MOI_gsb&8-}(e) and \ref{MOI_gsb&8-}(k)].

The dependence of seniority $\nu$ on the pairing gap reduction at $\hbar\omega=0$ calculated by the PNC-CSM method are
\begin{equation}
\begin{aligned}
  \frac{\tilde{\Delta}_n(\nu^{2}8^{-},\nu_n=2)-\tilde{\Delta}_n(\mathrm{GSB},\nu=0)}{\tilde{\Delta}_n(\mathrm{GSB},\nu=0)} & \approx 4.4\%,   \\
    \frac{\tilde{\Delta}_p(\pi^{2}8^{-},\nu_p=2)-\tilde{\Delta}_p(\mathrm{GSB},\nu=0)}{\tilde{\Delta}_p(\mathrm{GSB},\nu=0)} & \approx 6.1\%. 
\end{aligned}
\end{equation}

At the bandhead $\hbar\omega=0$, the seniority $\nu$ dependence of the neutron (proton) pairing gap is reduced by about 4.4\% (6.1\%) for the two-neutron $\nu^{2}8^{-}$ (two-proton $\pi^{2}8^{-}$) bands, which is due to the Pauli blocking of orbitals near the Fermi surface. This results in nearly 37\% and 13\% larger of $J^{(1)}$ for the two-neutron and two-proton $8^{-}$ bands compared with GSB in $^{252}$No [see Figs. \ref{MOI_gsb&8-}(e) and \ref{MOI_gsb&8-}(k)]. For the experimental $J^{(1)}$, the bandhead of the $8^{-}$ band is larger than the GSB by about $30\%$ in $^{252}$No. The main configuration of the $8^{-}$ band at the low frequency is assigned as the two-neutron $\nu 7/2^{+}[624] \otimes \nu 9/2^{-}[734]$ state. It is the $4.4\%$ pairing gap reduction for the two-neutron $\nu^{2}8^{-}$ configuration state contributes to the $\sim30\%$ increase of $J^{(1)}$ for the $8^{-}$ bands compared to the GSB. As for $^{244}$Pu and $^{250}$Fm, it is similar. The $5\%$ pairing gap reduction of the two-neutron $\nu^{2}8^{-}$ configuration state result in the $\sim25\%$ increase of the $J^{(1)}$ for the $8^{-}$ bands compared with their GSBs at the bandhead.

\section{Summary}{\label{Sec:summary}}
The $K^{\pi}=8^{-}$ isomeric states and rotational bands in the $N = 150$ isotones from $^{244}$Pu to $^{254}$Rf are investigated systematically by using the cranked shell model with the pairing correlations treated by the particle-number-conserving method. The experimental excitation energies of these isomers and moments of inertia of the ground-state bands and $K^{\pi}=8^{-}$ rotational bands are reproduced very well by the PNC-CSM calculations.\par 


The calculated two-neutron state with the configuration $\nu 7/2^{+}[624] \otimes \nu 9/2^{-}[734]$ is the lowest $8^{-}$ state for all the $K^{\pi}=8^{-}$ isomers studied in this work, which is the evidence of deformed neutron shell at $N=152$. As for proton, the present calculations give the $8^{-}$ state with two-proton $\pi 7/2^{-}[514] \otimes \pi 9/2^{+}[624]$ configuration at very low energy only for $^{252}$No and $^{254}$Rf due to the proton shell gap at $Z=100$.

There are upbendings of the two-neutron $8^{-}$ bands in the lighter isotones $^{244}$Pu, $^{246}$Cm, and $^{248}$Cf, while behaviors are much plainer for the bands in the heavier isotones $^{250}$Fm, $^{252}$No, and $^{254}$Rf. The upbending at frequency $\hbar\omega\approx 0.20$ MeV of $8^{-}$ band in $^{244}$Pu is mainly caused by the suddenly gained alignments from protons. Particularly, the interference term $j_{x} (\pi 7/2^{+}[633] \pi 5/2^{+}[642])$ give a considerable contribution to the suddenly increased alignment. In contrast, the proton Fermi surface of $^{250}$Fm locates just below the $Z=100$ deformed subshell, alignments gained from protons are quite gentle, there is no upbending of alignment of the $8^{-}$ band in $^{250}$Fm.\par

The irregularity of moment of inertia observed in the $K^{\pi}=8^{-}$ rotational band in $^{252}$No can be explained by the mixing of two-neutron $\nu 7/2^{+}[624] \otimes \nu 9/2^{-}[734]$ configuration dominating at low $\omega$ with two-proton $\pi 7/2^{-}[514] \otimes \pi 9/2^{+}[624]$ configuration dominating at high $\omega$.\par

The bandhead $J^{(1)}$ of the $8^{-}$ bands is larger than that of the ground-state bands by about $20\%-30\%$ in the $N=150$ isotones. This is attributed to the $\sim 5\%$ pairing gap reduction of the two-neutron $\nu 7/2^{+}[624] \otimes \nu 9/2^{-}[734]$ configuration state compared to the ground-state band at the bandhead. 

\section*{ACKNOWLEDGMENTS}
This work is supported by the National Natural Science Foundation of China (Grant Nos. 12475121 \& U2032138) and the National Key R$\&$D Program of China (Contract No. 2023YFA1606503).

\bibliography{references}
\bibliographystyle{apsrev4-1}
\end{document}